\documentclass{vgtc}                          

\graphicspath{{figures/}{pictures/}{images/}{./}} 

\usepackage{times}                     

\usepackage{tabu}                      
\usepackage{booktabs}                  
\usepackage{lipsum}                    
\usepackage{mwe}                       
\usepackage{hyperref}
\usepackage{hyperxmp}
\usepackage{graphicx}
\usepackage{subcaption}
\usepackage[all]{hypcap}

\usepackage{mathptmx}                  

\usepackage{multirow}
\usepackage{array}
\usepackage{amssymb}
\usepackage{amsmath}
\usepackage{listings}
\usepackage[dvipsnames]{xcolor}
\newcommand{\inlinehdr}[1]{\vspace{1.0ex}\noindent{\textbf{#1}}}

\onlineid{1017}

\vgtccategory{Research}

\ieeedoi{xx.xxxx/TVCG.201x.xxxxxxx}

\title{Managing Data for Scalable and Interactive Event Sequence Visualization}

\author{Sayef Azad Sakin\texorpdfstring{\thanks{email: sayefsakin@sci.utah.edu}}{email: sayefsakin@sci.utah.edu}%
\and Katherine E. Isaacs\texorpdfstring{\thanks{e-mail: kisaacs@sci.utah.edu}}{e-mail: kisaacs@sci.utah.edu}}%
\affiliation{\scriptsize SCI Institute and Kahlert School of Computing, The University of Utah}

\teaser{
    \centering
    \captionsetup[subfigure]{justification=centering}
    \begin{subfigure}[t]{0.49\columnwidth}
        \includegraphics[width=\textwidth]{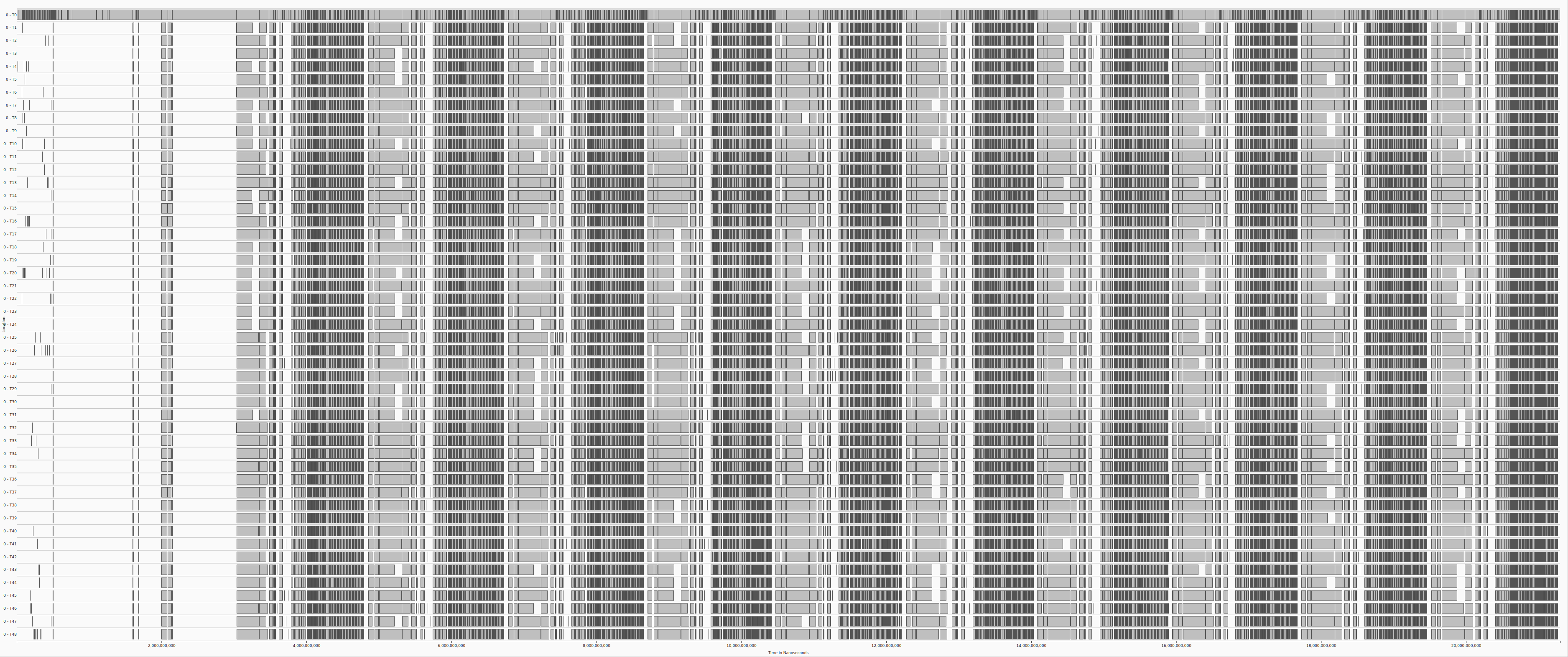}
        \caption{Raw Visualization, Horizontal Resolution = 3672, \\Data Fetch Time 499ms, Ground Truth}
    \end{subfigure}
    \begin{subfigure}[t]{0.49\columnwidth}
        \includegraphics[width=\textwidth]{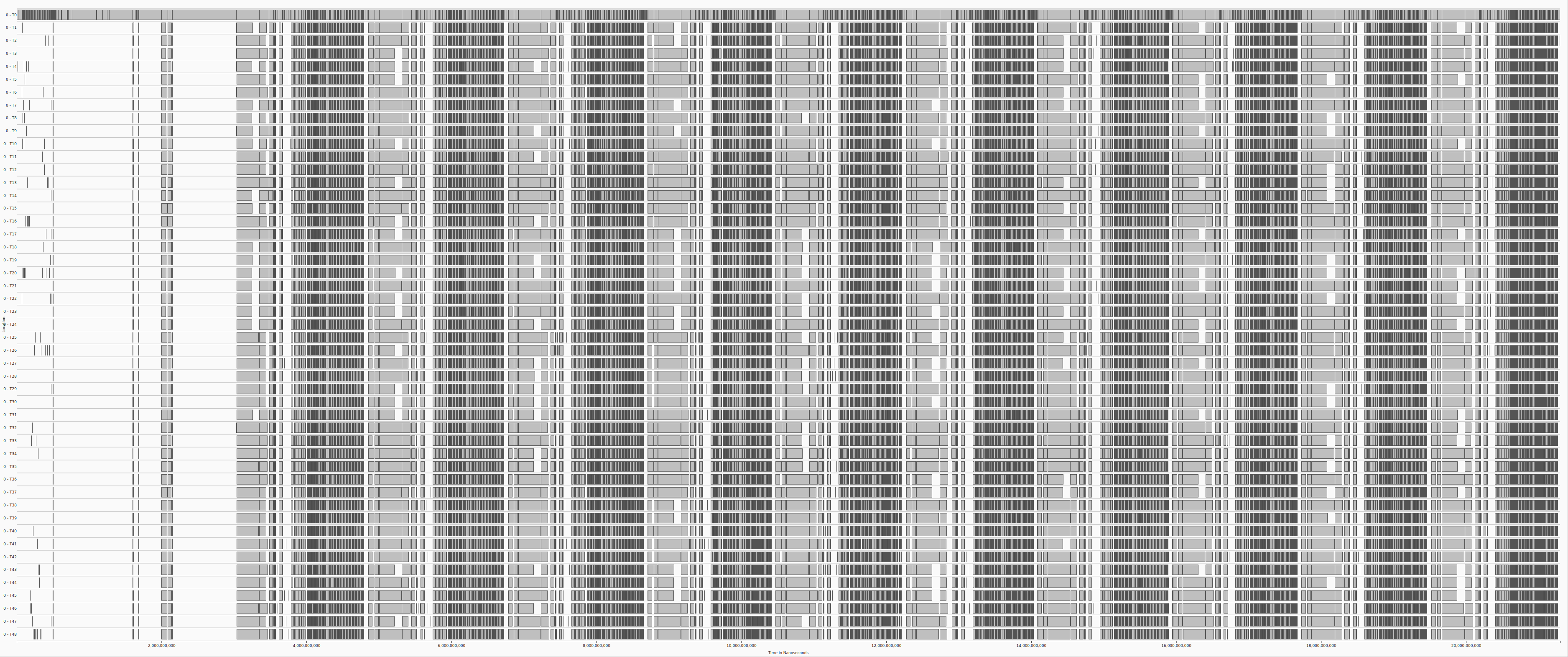}
        \caption{Summarized using ESeMan, Summarization window size = 1 pixel, \\Data Fetch Time 53ms, SSIM=1.00}
    \end{subfigure}
    \hfill
    \begin{subfigure}[t]{0.49\columnwidth}
        \includegraphics[width=\textwidth]{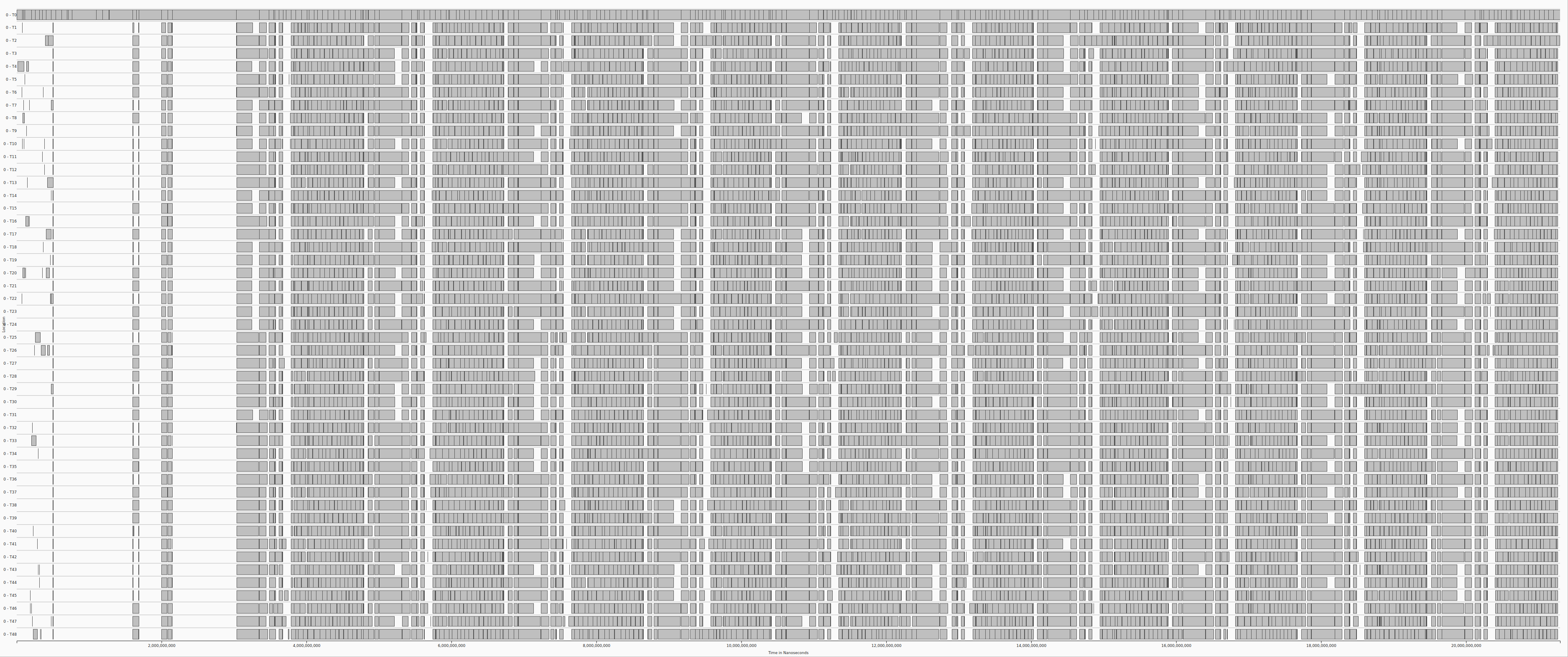}
        \caption{Summarized using ESeMan, Summarization window size = 16 pixels, \\Data Fetch Time 22ms, SSIM=0.74}
    \end{subfigure}
    \hfill
    \begin{subfigure}[t]{0.49\columnwidth}
        \includegraphics[width=\textwidth]{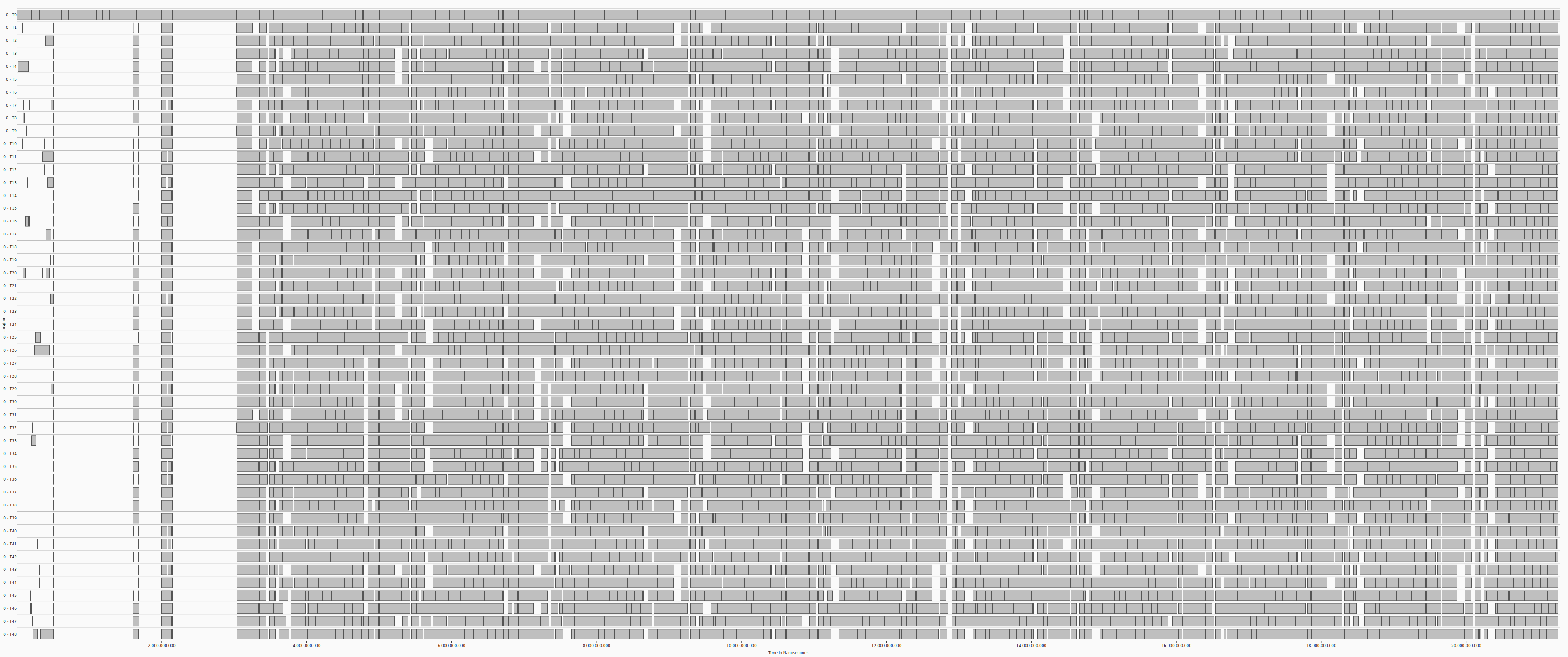}
        \caption{Summarized using ESeMan, Summarization window size = 32 pixels, \\Data Fetch Time 3ms, SSIM=0.67}
    \end{subfigure}
    \caption{Parallel timeline charts constructed from 160K events distributed across 49 processors (rows) of a parallel program. Individual gray boxes with darker borders represent single events. The upper left (a) shows a naive rendering of all of the data, while (b - d) show renderings from data fetches against our ESeMan solution configured with different levels of summarization. The naive solution requires 499 milliseconds (ms) for the data fetch, which is outside best practices for interactivity. Our ESeMan solution retrieves data for an equivalent chart in one tenth the time. Furthermore, ESeMan can be tuned to trade off accuracy for fetch time. For example, in (c), the summarization window is increased to 16 pixels, resulting in noticeable differences in the visualization (quantified by the SSIM score), but half the fetch time. We go further in (d), increasing the summarization window to 32 pixels for a significant decrease in data fetch time with further degradation to accuracy.}
    \label{fig:teasers}
}

\abstract{
Parallel event sequences, such as those collected in program execution traces and automated manufacturing pipelines, are typically visualized as interactive parallel timelines. As the dataset size grows, these charts frequently experience lag during common interactions such as zooming, panning, and filtering. Summarization approaches can improve interaction performance, but at the cost of accuracy in representation. To address this challenge, we introduce ESeMan (Event Sequence Manager), an event sequence management system designed to support interactive rendering of timeline visualizations with tunable accuracy. ESeMan employs hierarchical data structures and intelligent caching to provide visualizations with only the data necessary to generate accurate summarizations with significantly reduced data fetch time. We evaluate ESeMan's query times against summed area tables, M4 aggregation, and statistical sub-sampling on a variety of program execution traces. Our results demonstrate ESeMan provides better performance, achieving sub-100ms fetch times while maintaining visualization accuracy at the pixel level. We further present our benchmarking harness, enabling future performance evaluations for event sequence visualization.
}


\begin{document}

\firstsection{Introduction}

\maketitle

Event sequence data is prevalent in domains like healthcare, manufacturing, and computer program monitoring where they are frequently analyzed for event progression, relative durations of events, and similarities between sequences in time. For example, in computing, execution traces are sequences of events describing computing activity per hardware resource and are used in identifying performance bottlenecks and assessing optimization of resource use~\cite{isaacs2014state, davidson2023qualitative}. In healthcare, patient timelines encode clinical events such as diagnoses, treatments, and medication administration; analyzing these sequences can inform personalized care decisions and improve treatment outcomes~\cite{antweiler2022uncovering, bernard2018using}. In manufacturing, managers track sequences of operations across distributed facilities to monitor progress, detect delays, and coordinate workflows~\cite{lee2022linear, undozerov2021spring}. Across all these domains, effective analysis of event sequences enables domain experts to extract actionable insights from complex, time-evolving systems.

Several of these domains can produce datasets with millions of events. In the computing case, datasets grow as the execution time and number of parallel resources increases. Manufacturing similarly can exhibit lengthy time span broken into sequences by large numbers of machines. As the number of events increases, so does the computational burden of querying, aggregating, and redrawing them. Therefore, these datasets can tax interactive visual representations of this data, leading to significant lag. For \textbf{parallel timeline charts}, a widely-use visual idiom for event sequences and a key component of a Gantt chart, this lag can limit common exploratory operations such as panning and zooming.

This problem generally arises in domains where automated event logs are generated over time. For example, during software performance analysis using event traces, a timestamped log of software in execution, programs running for a longer time on a higher number of computing resources generally create execution traces with large numbers of events~\cite{isaacs2014state}. While execution traces can theoretically grow to hundreds of millions or even billions of events, such massive traces are generally impractical for visual analysis. Similarly, in healthcare systems, patient monitoring devices can generate a large stream of time-stamped health indicators or biometric readings~\cite{bai2024global}. In practice, for visual exploratory analysis, analysts extract a specific and manageable set of events, such as events around certain I/O operations in high performance computing (HPC) systems~\cite{davidson2023qualitative}, or a harmful period of certain disease spread~\cite{hirakata2022exploring} in healthcare. A targeted and manageable event subset for interactive analysis typically contains thousands to millions of events. This selective strategy helps reduce both system overhead during data collection and improves analysts' experience in comprehension and visual exploration.

However, even with reduced data size, visualizing and interacting with millions of events in a parallel timeline still results in obstructive \textit{visual latency}, the delay between when a user initiates an interaction (such as zooming, panning, or filtering) and when the system completes rendering the updated view. For exploratory visual analysis, latency should be less than 100 milliseconds (ms) to achieve interactivity~\cite{battle2019role}. When latency exceeds the 100 ms threshold, it disrupts users' iterative analytical workflow and reduces the ability to explore relationships, test hypotheses, and gain insights from the event data~\cite{liu2014latency, battle2020database, battle2019role}.

Balancing interactivity while maintaining an accurate representation remains a significant challenge in visualizing large event sequences~\cite{guo2021survey, yeshchenko2024survey}, requiring careful coordination of data management throughout the entire visualization pipeline. Managing the data for visualization includes identifying user-driven queries, fetching relevant data from storage, passing it efficiently between components (disk, memory, cache), processing it based on interaction context, and preparing it for rendering. Poor data management, such as fetching irrelevant data, redundant data transfers, format mismatches, or lack of reuse—can significantly increase both latency and memory consumption. These inefficiencies accumulate across the pipeline and undermine the overall responsiveness and usability of the system. 

To address the need for interactivity in parallel timeline charts with millions of events, we introduce \texttt{ESeMan (Event SEquence MANager)}, a data management solution for efficient data fetching for parallel timeline charts. ESeMan uses an internal hierarchical spatial index that produces event summaries matching the resolution of the visualization, thereby avoiding querying individual events less than a pixel wide. This approach provides near pixel-perfect accuracy while greatly reducing data fetch time. Furthermore, by tuning the fidelity requested, ESeMan can produce more aggressive summaries, allowing users to tune the balance between accuracy and latency.

We compare ESeMan to common data management strategies for efficient querying of interactive charts, including in-memory databases, summed area tables, statistical sub-sampling, and M4 optimization. ESeMan outperformed the alternatives, achieving sub-100 ms fetch times for datasets with millions of events while maintaining near perfect similarity when compared to the non-summarized parallel timeline chart. We also show the trade off between fetch time and accuracy when we vary the fidelity. 

Our paper is organized as follows. 
We first describe terminology related to event sequence data and parallel timeline charts along with related work (\autoref{sec:background}). Then we describe our proposed data management solution (\autoref{sec:solution}) and our evaluation setup and results (\autoref{sec:eval}). Finally we discuss recommendations for configuring ESeMan and conclude.

\section{Background and Related Work}
\label{sec:background}

We first present background and terminology for parallel timeline charts. We then discuss related work regarding handling large-scale data for parallel timeline charts, including perspectives on the use of databases, summarization and reduction techniques, and existing open-source and commercial tools.

\begin{figure}[t]
  \centering
  \includegraphics[width=1.0\columnwidth]{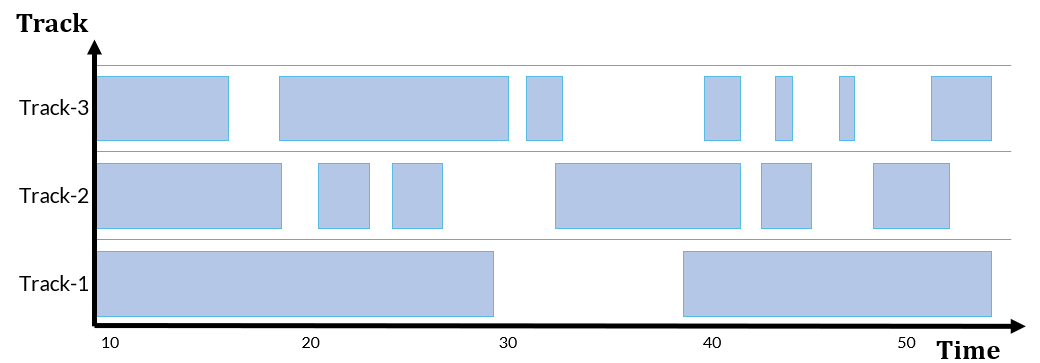}
  \caption{Parallel event sequences on three tracks and a time window are visualized using a parallel timeline chart.}
  \label{fig:gantt_desc}
\end{figure}

\subsection{Parallel timeline charts}
A parallel timeline chart is a visualization idiom representing multiple related sequences of events. One axis represents time while the other is used to represent track. Each event can be represented as a bar in its assigned track that spans its duration on the time axis. 
These charts are commonly used in domains such as software execution traces~\cite{isaacs2014state, ezzati2017multi}, user activity logs~\cite{wilson2003gantt}, manufacturing system monitoring~\cite{jo2014livegantt}, and medical workflows~\cite{antweiler2022uncovering}. 
Typically in these applications, events are defined their \textit{start time} and \textit{end time}. Events may have additional data in the form of attribute-value pairs. Typically, the multiple sequences are defined by partitioning the total set of events by one of these attributes, e.g., a computational resource in a parallel computing trace or a patient in a set of patient histories. We refer to the assigned sequence as a \textbf{track}, following the terminology of Sakin et al.~\cite{sakin2024gantttaxonomy}.

Depending on the time span and track span of the data, panning and zooming in the parallel timeline chart may be required to keep the events larger than a pixel in width or height. \autoref{fig:gantt_desc}  shows an example of a parallel timeline chart zoomed to show events in three tracks. Some bars are only partially shown at the given zoom level. 

There are several ways a parallel timeline chart may choose to render sub-pixel events---including drawing them as single pixel events, omitting them entirely, or aggregating them. Aggregating sub-pixel events is a form of event \texttt{summarization}~\cite{guo2021survey}. An \textbf{event summary} is a reduced representation of one or more event sequences to support the \texttt{summarization} analysis task.

Interactivity and aggregation for parallel timeline charts have been a long-standing problem. Davidson et al.~\cite{davidson2023qualitative} conducted an interview study on different distributed trace analysis tools, noting that maintaining data quality and obtaining various levels of aggregation results are one of the primary problems. Literature surveys in software visualization~\cite{isaacs2014state, ezzati2017multi} further highlight the importance of managing data on varying scales. Our data management solution attempts to address challenges in maintaining data quality while giving options for summarization at various scale levels.

\subsection{Databases in Event Sequence Visualization}
Database management systems (DBMS) for visualizing event sequence data have been explored in several works~\cite{bell2003paraprof, antweiler2022uncovering, sun2021daisen, battle2016dynamic, agarwal2013blinkdb, heer2023mosaic, antweiler2022uncovering}, but do not optimize for changing representations for summarization. Distributed databases have been used for handling large software performance data~\cite{moritz2015perfopticon, kaldor2017canopy, kruchten2022vegafusion, shen2023qevis}, including to handle streaming cases~\cite{kesavan2020visual, khan2023web}. Distributed databases are heavily dependent on fast network communication in a distributed environment, query optimization, distributed query execution, caching, and progressive rendering. These solutions are limited by the latency due to network communication, often resulting in total latencies over the recommended threshold for interactivity~\cite{kaldor2017canopy, kruchten2022vegafusion}. Furthermore, they also exhibit configuration challenges for users. We aim to provide a solution that provides latency within the interactivity threshold limit without the burden of network configuration for the visualization designers.

Battle et al.~\cite{battle2020database} presented a database benchmarking study focusing on real-time interactive queries of large data. Their results significantly highlight the limitations of different database solutions in maintaining interactivity with a threshold of 100ms. In the study, DuckDB was noted as one of the best-performing databases, which motivated us to use and compare DuckDB in our evaluation conditions.

A structured review of hybrid data management  solutions that combine data structures and DBMSs was presented by Battle et al.~\cite{battle2020structured}. By their categorization, our proposed data management solution would be considered an \texttt{Indexing Technique} because we use a KD-Tree for indexing and the Lightning Memory-Mapped Database Manager (LMDB)~\cite{lmdb} for its querying. The review also described Data Sketching~\cite{cormode2017data, budiu2015interacting} and statistical sub-sampling-based techniques~\cite{budiu2019hillview, de2010zinsight, drebes2014aftermath} to achieve fast query performance with guaranteed error bounds. Data sketching and statistical subsampling are practical for streaming data to provide summarization within controlled errors. However, the memory requirement for the indexed data using these approaches increases proportionally with the data size. Therefore, these approaches can fall short in scenarios for retrieving exact summaries during post-hoc analysis. We include comparisons to a representative subsampling approach in our evaluation.

\subsection{Summarization and Reduction for Charting}
Data summarization has been implemented in visualization grammars~\cite{satyanarayan2016vega, heer2023mosaic, mcnutt2022no} to reduce fetched data size by optimizing queries and caching prior query results on materialized views~\cite{battle2020structured}, which is a snapshot of the prior query results. Generally, after optimizing the query, a summarized version of the prior results is stored in the database-specific materialized views.
Storing materialized views requires additional memory on the visualization side. A visualization grammar for event sequences is absent in these works. Furthermore, grammar-based visualization tools~\cite{satyanarayan2016vega, heer2023mosaic} utilize M4 optimization~\cite{jugel2014m4} to retrieve the data, which is not optimized for event sequence data. 

Without using a database, linear data structures can be used to generate an aggregated summary, also trading off accuracy and speed. In data visualization, summed area tables~\cite{sakin2022traveler} and largest triangle three buckets (LTTB)~\cite{van2022plotly} have been used KD-box~\cite{zhao2021kd} and Kyrix-J~\cite{tao2020kyrix} use KD-Tree in summarizing time series data visualization. Our proposed solution applies KD-Trees for event sequence data.

Neural networks~\cite{yeshchenko2021visual, tao2022kyrix, krawczuk2021anomaly} and genetic algorithms~\cite{jo2014livegantt} have been used to generate event sequence summaries highlighting domain-specific  anomalies. Clustering solutions like model clustering~\cite{nesi2023summarizing}, clustering for large applications~\cite{isaacs2014combing}, and agglomerative clustering~\cite{pasupathi2021trend} have been used for trend analysis for time series data. While clustering approaches are generally time-consuming and sensitive to the distribution of the data, we implemented an agglomerative clustering approach as an option in our solution, comparing its results to our tree-based option.

\subsection{Commercial event sequence visualizations}
Several open-source tools offer event sequence analysis in various capacities. Google's Perfetto~\cite{Perfetto} trace analyzer uses the web browser's in-memory database. Rather than summarize sub-pixel events, it samples them, thus dropping data from the display. Our solution provides similar response times to Perfetto with a summarization approach. 

Grafana~\cite{grafana} provides a state timeline to visualize event sequence data. It sub-samples data through a pagination scheme. Drawing an overall overview or summary views on different scale levels is not possible in Grafana. Additionally, it has limited functionality to de-clutter event data with a high frequency of smaller events. In our proposed solution, we primarily focus on generating summaries at various granular levels without decreasing data accuracy.

Commercially available solutions~\cite{Tableau, Jaeger, Zipkin, AmazonXRay, Datadog, Lightstep, Flurry, NewRelic, Honeycomb, Elastic, powerbi} serve diverse market segments and specific user needs. All of these solutions require substantial infrastructure investments, proprietary cloud backends, and vendor lock-in that can limit accessibility and flexibility. These platforms, while powerful, create barriers to entry for researchers, small organizations, and developers who need precise control over their analysis workflows without the overhead of sophisticated storage backends or licensing costs. Our solution aims to remove the barriers by providing an open-source solution that delivers both accuracy and performance while remaining accessible to researchers, developers, and analysts.

\section{ESeMan Library for Interactive Event Queries}
\label{sec:solution}

We present our library, ESeMan (\textit{Event SEquence MANager}), for supporting interactive querying of event sequence data for rendering parallel timeline charts. ESeMan assumes a \textit{data summarization} approach where sub-pixel events will be aggregated. Users of ESeMan can also select a larger summarization window, trading off pixel-level accuracy for lower latency.

We first provide an overview of the approach. Then we discuss the details of the hierarchical spatial index used by ESeMan to represents and the compressed representation of event attributes. Finally, we discuss how the fetch is done in the tree and caching mechanisms used to further accelerate the query.

\subsection{ESeMan Overview}

\begin{figure}[t]
  \centering
  \includegraphics[width=\columnwidth]{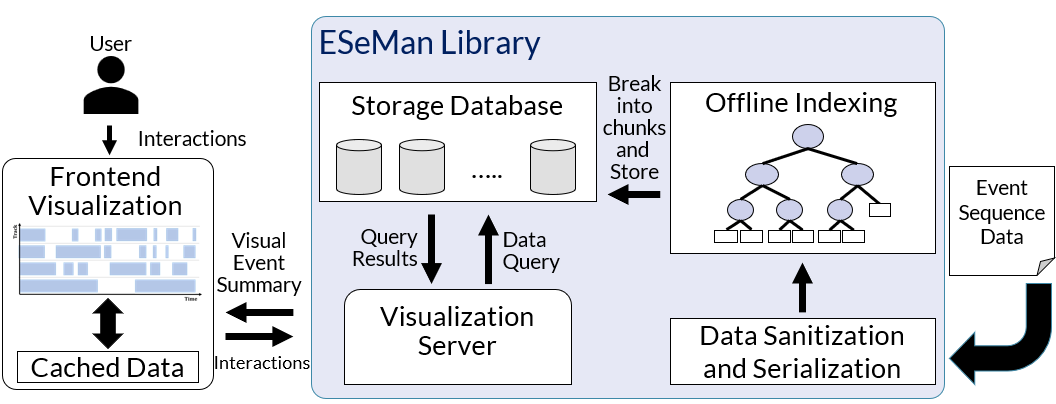}
  \caption{ESeMan provides support for low latency interactions in parallel timeline charts through fast querying. It ingests and cleans event sequence data, building indices on the events and then serving summarized events to match data queries. These summarized events provide pixel accuracy with less data sent. Stronger summarization can trade accuracy for latency.}
  \label{fig:pipeline}
\end{figure}

We assume a pipeline (\autoref{fig:pipeline}) in which visualization authors use ESeMan to store, process, and query their event sequence data towards drawing their own parallel timeline chart. Specifically, ESeMan supports parallel timeline charts where short events (i.e., sub-pixel events) are summarized. When the designer's visualization needs data, they send the time and track spans to ESeMan, along with any filtering conditions, and are sent back the summarized data for rendering.

When given a new dataset, ESeMan first preprocesses the data. Data is sanitized to remove unnecessary data and serialized into ESeMan's internal format. ESeMan then performs hierarchical spatial indexing (\autoref{subsec:hierarchy}) to generate event summaries which are stored in the storage server(s). An internal visualization server handles queries from the visualization, communicating with the storage server to fetch the relevant summaries. 

By default, ESeMan assumes the visualization is requesting pixel-level summaries which would provide high accuracy compared to a raw depiction that drew each event. Users can request coarser summaries by changing the pixel window by which the summaries are fetched, resulting in lower latency due to less data to transmit and a shallower query against the hierarchical index.

\subsection{Hierarchical Spatial Indexing in ESeMan}
\label{subsec:hierarchy}

We build a hierarchical structure of summarizations of event sequence data where leaves represent the individual durational events and the root represents all events in the tree. Internal node levels represent different partitions of the parallel timeline space. Given a time span, track span, and measure of accuracy (e.g., relation between pixels and time and track), a data query need only search from the root of structure to the required accuracy, thereby speeding up the fetch time. 

We implemented three different approaches for the hierarchical data structure, two which create a forest of trees, one per track, and one which creates a single tree across all time and tracks. The forest approach serves data where the ordering of tracks is expected to be dynamic, such as when re-order or per-track filter operations are frequently used. The single tree approach serves cases where these interactions aren't present, such as when a fixed track order is highly meaningful or when such interactions aren't implemented. We offer KD-tree~\cite{bentley1975multidimensional} approaches in both the forest and single-tree cases, with an additional agglomerative clustering approach in the forest case. We describe each approach below and analyze their relative efficiency in \autoref{sec:eval}.

\begin{figure*}[t]
    \centering
    \begin{subfigure}[t]{0.9\columnwidth}
        \includegraphics[width=\textwidth]{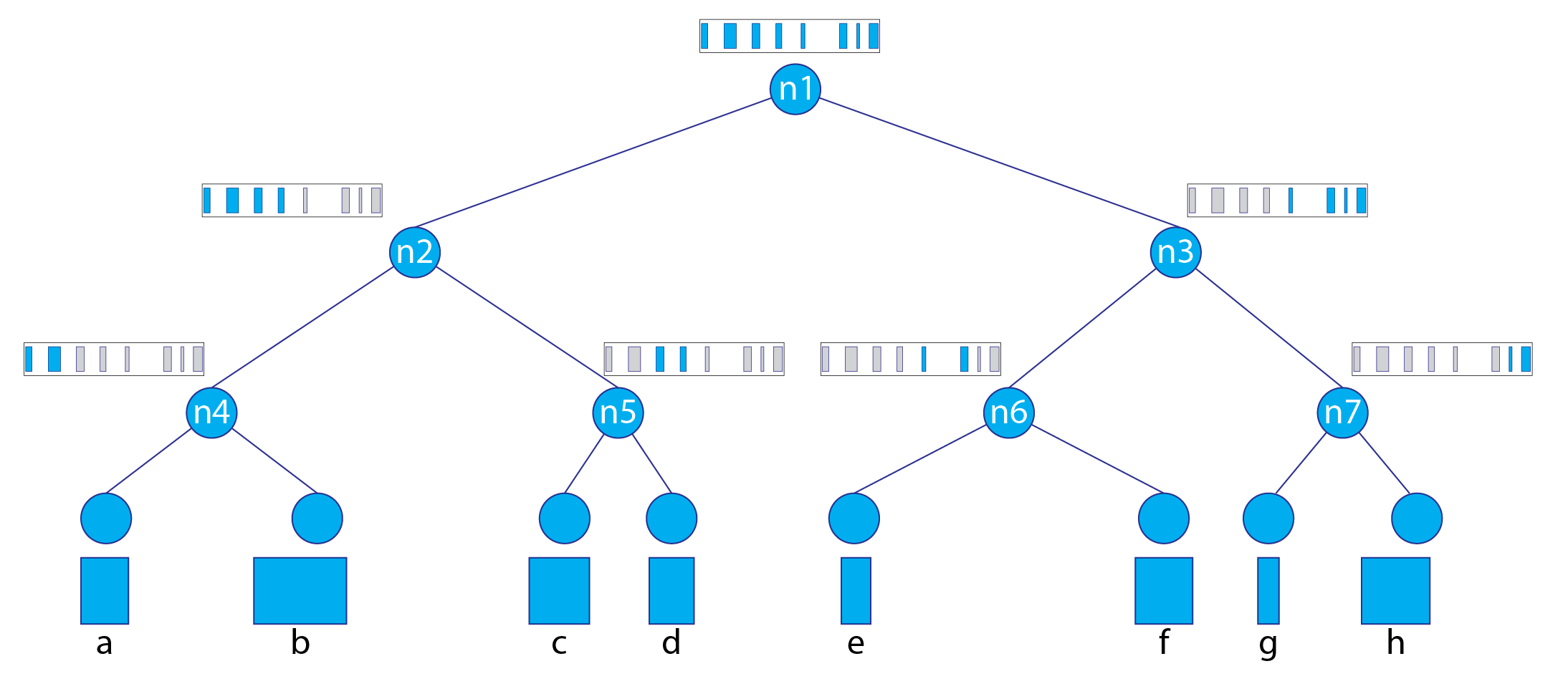}
        \caption{Storing events using one-dimensional KD-Tree}
    \end{subfigure}
    \hfill
    \begin{subfigure}[t]{0.9\columnwidth}
        \includegraphics[width=\textwidth]{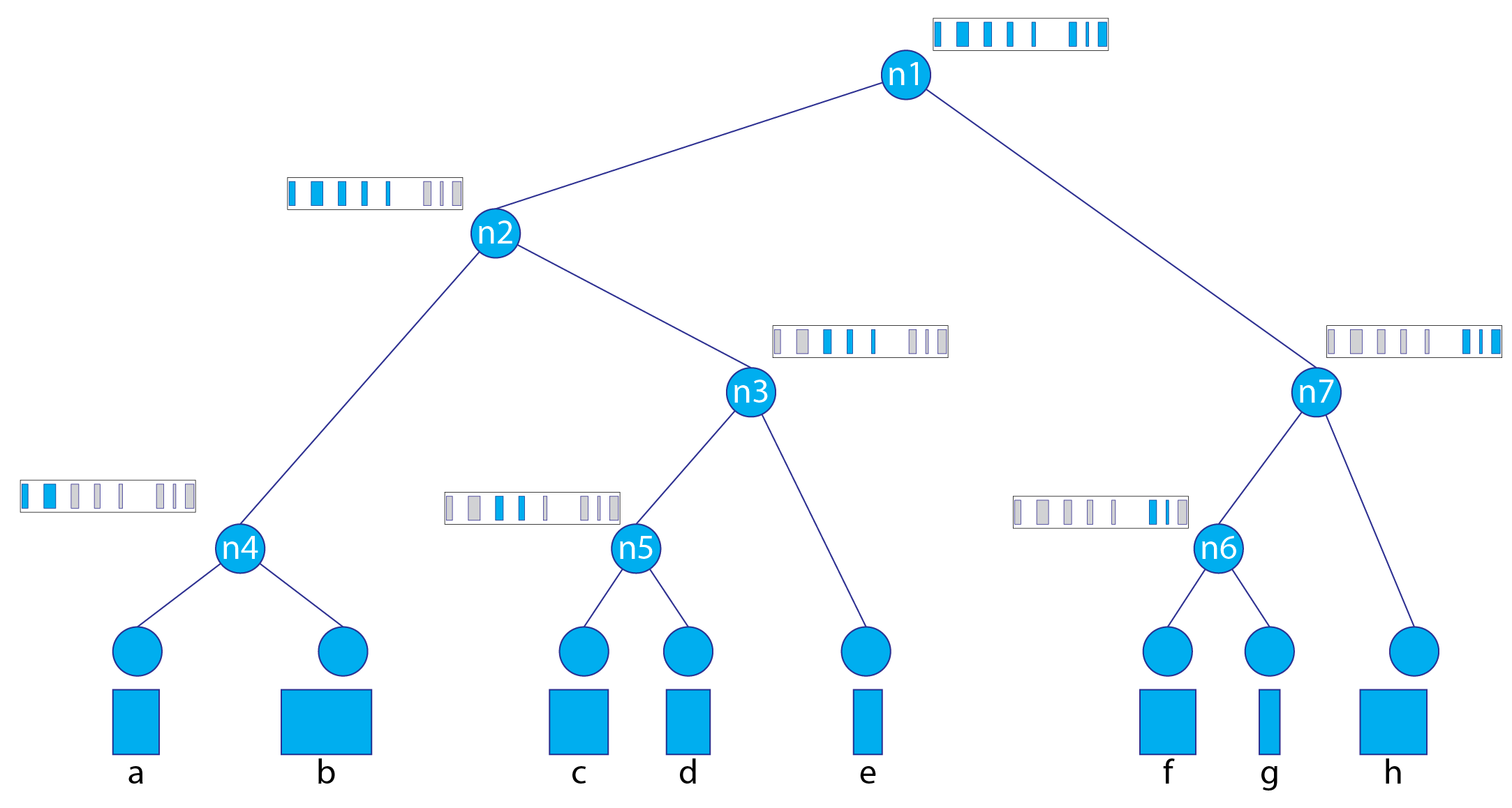}
        \caption{Storing events based on agglomerative clustering}
    \end{subfigure}
    \caption{Storing events (blue boxes at the leaf) of a single track using a hierarchical structure, based on (a) a one-dimensional KD-Tree, and (b) agglomerative clustering. Intermediate nodes $n2-n7$ store a list of  events (highlighted with small blue boxes beside each node). The root node $n1$ represents all the event sequences of a single track. During querying, when a pixel is represented by a node with multiple events, the temporal bounds of the node's event list are sent rather than the individual events, resulting in a summary depiction.}
    \label{fig:hierarchy}
\end{figure*}

\begin{figure*}[t]
  \centering
  \includegraphics[width=0.8\textwidth]{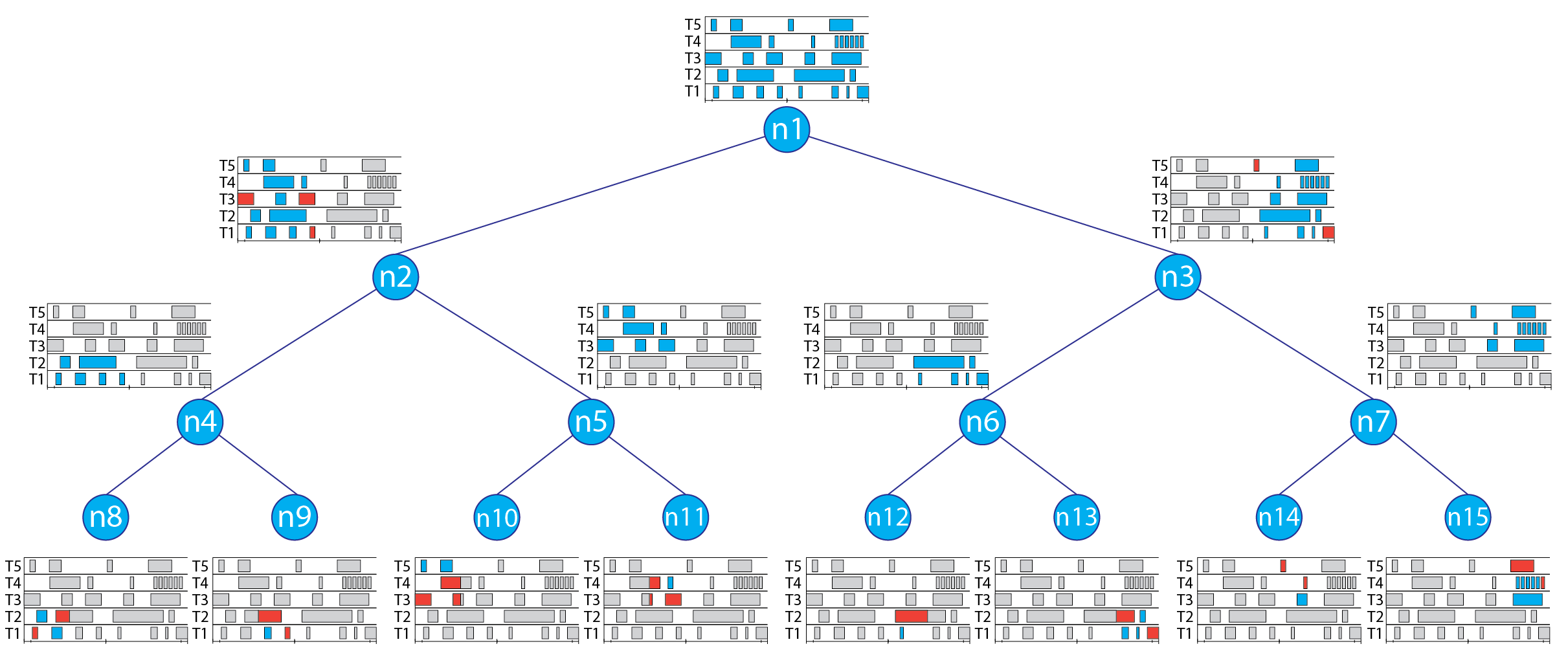}
  \caption{Event sequences distributed across five tracks ($T1$ to $T5$) are stored using a two-dimensional KD-Tree. Intermediate nodes from $n2$ to $n15$ store the summarized events (highlighted with small red and blue boxes beside each node where red indicates events on the node's temporal boundary). The root node $n1$ stores all the event sequences across five tracks. During querying, when a pixel is represented by a node with multiple events, the temporal bounds of the node's event list are sent rather than the individual events, resulting in a summary depiction.}
  \label{fig:twodkdt}
\end{figure*}

\inlinehdr{One-dimensional KD-Tree to summarize events per track.}
We construct the tree in a top-down fashion. The root contains the summary of all the events along a track. Iteratively, nodes representing a time-span with more than two events are split into child nodes based on the temporal order of the events contained. The policy that dictates how the events are divided at each node is referred to as the \emph{splitting rule}~\cite{gaede1998multidimensional, bentley1975multidimensional}. We follow a \emph{fair splitting rule}, which splits nodes such their children have an equal number of events. When there are an odd number of events, the extra is given to the right sub-child. In our preliminary testing, when compared to other candidate splitting rules, this rule resulted in the highest accuracy representations when summarization is set to pixel-size. \autoref{fig:hierarchy}(a) show which events are summarized by which nodes in the one-dimensional KD-tree.

\inlinehdr{Two-dimensional KD-Tree to summarize all events.}
We use a two-dimensional implementation of a KD-tree to summarize across the dimensions of \textit{time} and \textit{tracks} simultaneously. Like the 1D case, the tree is constructed in a top-down fashion, and the root node contains the summary of the entire parallel timeline chart. During splitting, a KD-Tree alternates between the horizontal (time) and vertical (track) dimensions. We follow the \emph{midpoint rule} while splitting for both dimensions to construct a balanced tree structure due to its high accuracy during preliminary testing versus other rules. 

When splitting in the time axis, we choose the midpoint of the time span bounded by the start of the earliest event and end of the latest event in the parent. If the midpoint intersects an event, it is split between the children. If it does not overlap with any event (across all tracks in the node-to-be-split), we adjust the time represented by the child nodes to snap to the events contained. When splitting in the track axis, we evenly divide the tracks. When there are an odd number of tracks, the extra track is assigned to the top region. Once a node contains only a single track, we switch from the midpoint rule to the fair splitting rule used in our 1D case.

\autoref{fig:twodkdt} shows how nodes partition events in a multi-track timeline chart in our two-dimensional KD-Tree. We highlight in red the events defining region boundaries when the midpoint does not intersect with an event.

\inlinehdr{Agglomerative clustering to summarize events.}
Agglomerative clustering is a bottom-up hierarchical clustering method that begins with each item as an individual cluster and iteratively merges the closest pairs based on a distance metric. As a clustering method, this technique is frequently used to help identify natural groupings within data~\cite{ackermann2014analysis}.

We define the distance between two events as the temporal gap between them---from the end of the earlier event to the beginning of the later one. We define the distance between groups of events using \textit{single linkage}, i.e., the smallest distance between any pair of events from the two clusters. This approach favors connecting temporally proximate events, resulting in compact clusters along the timeline and simplifying calculation of cluster distances. Each cluster is bounded by the earliest start time and latest end time of its constituent events, which is then used to generate summary representations. \autoref{fig:hierarchy}(b) illustrates this process.

\subsection{Summarizing Event Attribute Data}
\label{subsec:attributes}
We have discussed how we index and store the temporal data defining events. However, attributes associated with those events are frequently also queried by the visualization, e.g., for labels, coloring, and filtering. Therefore, after building the trees, we make a bottom-up pass to populate hash tables with attribute data within the summaries described by the internal tree nodes. The attribute values contained in the summary are made available but as the individual events are summarized, the individual attribute-value-too-event mapping is not.

For categorical attributes, we store the set of unique values observed in its subtree. The hash key is the name of the attribute and the value is a list of all values within the summary. This approach is fast for lookup and merging. We considered bitmaps for a more memory-efficient implementation, but these are limited to a fix number of values and require sending a full decoding key to the client.

As an example, in software execution trace data, an attribute named ``function'' holds program function names. If two child nodes have events with functions (\texttt{foo}, \texttt{bar}) and (\texttt{baz}) respectively, the parent node will store the list (\texttt{foo}, \texttt{bar}, \texttt{baz}) in its hash table.

We do not evaluate numerical attributes. However, we can apply an aggregator function during merging to also store these in hash tables. Each node would maintain one or more summary statistics (e.g., \textit{min}, \textit{max}, \textit{mean}, \textit{median}, \textit{standard deviation}) derived from the numerical values in its children. For instance, if one child node has an average duration of five seconds and another child has seven seconds, the parent node stores a simple average value of six seconds in its hash table. This propagation strategy enables each node in the hierarchy to provide high-level attribute summaries without storing the full event attribute values, ensuring scalability and faster accessibility of the attribute values.

\subsection{Fetching Data: Tree Traversal and Caching}
\label{subsec:caching}
When retrieving data, ESeMan traverses the hierarchy only to the depth necessary as determined by the time and track ranges given in the query, the number of pixels used by the visualization, and the accuracy requested in terms of pixels. To further aid efficiency, ESeMan also implements a caching mechanism to load and unload tree nodes based on these factors. We describe these strategies below.

Accuracy is controlled by the \textbf{pixel window} setting in ESeMan. By default, this is set to 1 pixel, meaning the summarizations are returned at the pixel level and thus we need to query internal nodes best representing every pixel. By selecting a larger pixel, the number of nodes needed is decreased, meaning we need to seek fewer nodes, transfer less data. Therefore, a larger pixel window will decrease data fetch time and visualization latency, but aggregate data across more pixels.

The tree traversal begins at the root and proceeds recursively down the tree. For each node, we check whether the node's time span overlaps with the query window. If the node lies entirely outside the range, it is discarded. If the node overlaps and its time span is larger than the time span represented by a single pixel window, we continue the traversal into its children. If the node's span is equal to or smaller than the time span represented by a single pixel window, the traversal terminates at that node. The time and track bounds of the node are sent to be visualized rather than having to send all of the events within the ndoe. When the node contains a single event, these bounds are equivalent.

To reduce data fetching from the storage server, we create a cache in the visualization server. The cache contains nodes that were traversed on the previous range-based query. During each successive range-based query, the server first consults the cache. If the node is found in the in-memory cache, its summary is immediately available for use. If the node is not cached, it is loaded from the storage server. We discard nodes from the cache that are not traversed during the current range query. This cache is designed to increase the performance of queries with a similar range, which may occur due to panning, zooming, or filtering, but will not aid in performance for random access queries of the parallel timeline chart.

\section{Evaluation}
\label{sec:eval}

We evaluate the performance and practicality of ESeMan through (a) a series of performance experiments comparing ESeMan to four alternative data management techniques frequently used in data visualization (below), a demonstration of ESeMan under different pixel window configurations (\autoref{subsec:tradeoff}), and (c) a qualitative comparison to Perfetto (\autoref{subsec:perfetto}). In these evaluations, we use a variety of execution trace datasets, representing computations on parallel clusters, for our tests because these datasets are frequently visualized using parallel timeline charts~\cite{isaacs2014state, sakin2024gantttaxonomy} and exhibit a variety of event sizes and distributions.

\subsection{Experimental Conditions}

To conduct the experiments, we extended the Traveler~\cite{sakin2022traveler} trace visualization tool with a benchmarking harness. Traveler renders parallel timelines in a single HTML5 \texttt{Canvas}. During the experiments, the horizontal resolution of the visualization canvas was set at 3672 pixels, fitting in an ultra high definition display with room for chart axes.
We implemented seven data management configurations (three of ours, four alternatives) within the Traveler benchmarking ecosystem, each representing a distinct aggregation and storage strategy. These configurations represent data optimization approaches commonly used in interactive visualization systems:
\begin{itemize}
    \itemsep0em 
    \item \textbf{ESeMan-1DKDT} (with LMDB~\cite{lmdb}) – our solution using a one-dimensional KDT per track over a memory-mapped key-value storage (LMDB) optimized for high-throughput reads.
    \item \textbf{ESeMan-KDT} (with LMDB) – our solution using a two-dimensional KDT, also over a memory-mapped key-value storage optimized for high-throughput reads.
    \item \textbf{ESeMan-Agg} (with in-memory Agglomerative Clustering~\cite{mullner2013fastcluster}) – our solution using a hierarchical in-memory clustering method per track.
    \item \textbf{Summed Area Table}~\cite{crow1984summed} (with DiskCache~\cite{diskcache} over MySQL) – precomputed aggregations with persistent caching, used by Traveler~\cite{sakin2022traveler}. Represents a linear data structure.
    \item \textbf{Naive} (with DuckDB~\cite{raasveldt2019duckdb}) –  Basic range-based SQL queries directly over flat trace data. Represents a baseline use of a relational database.
    \item \textbf{Statistical Sub-sampling} (with DuckDB) – data reduction via reservoir sampling\cite{duckdb-reservoir-sampling}. Represents a sub-sampling approach. 
\item \textbf{M4 Optimization}~\cite{jugel2014m4} (with DuckDB) – uses time-aware approximation and represents an approach employed by widely-used visualization libraries such as Vega\cite{vega}.
\end{itemize}

The ESeMan configurations (\textbf{ESeMan-1DKDT, ESeMan-KDT, and ESeMan-Agg}) are implemented using our proposed hierarchical spatial indexing. In these implementations, the KD-Tree based configurations use the same Lightning Memory-Mapped Database Manager (LMDB)~\cite{lmdb}. LMDB is a lightweight, memory-mapped key-value store optimized for fast, read-heavy workloads. It is more suitable for fast retrieval of structured key-value based data that our solution utilizes. Thus, we chose it over a relational database which does not match ESeMan's KD-tree workload needs.
 
We include the \textbf{Summed Area Table} configuration as a representative of a linear data structure~\cite{crow1984summed} to assess the differences between linear and hierarchical~\cite{gaede1998multidimensional} data structures. This approach has already been used for parallel timeline-based data visualization solutions~\cite{sakin2022traveler}.

\textbf{DuckDB} was chosen for relational database use for its high-performance in-memory query engine that supports SQL-style range queries without requiring external infrastructure. This makes it a suitable baseline for evaluating both naive and approximate querying methods. Among the three DuckDB-based configurations, we chose to compare against the \textbf{M4 Optimization} technique, a proven~\cite{jugel2014m4} time-series summarization technique used by scalable charting solutions like Vega~\cite{vega} and Mosaic~\cite{heer2023mosaic}. By using LMDB and DuckDB across our configurations, we attempt to isolate the effects of data structures and summarization strategies while keeping the storage layer fast and consistent with interactive performance targets. 

\subsection{Datasets}
We compare each condition across six distinct datasets, which vary in size and event characteristics, as described in \autoref{tab:datasets}. All datasets are execution traces of applications written in the HPX~\cite{Kaiser2020HPX} runtime, which have been previously visualized by the visualization community~\cite{sakin2022traveler}. The datasets feature different densities, durations, and clusterings of events. All available tracks were loaded in lexicographic order by their hardware thread ID.

All but the ``Synthesized'' dataset were collected using APEX~\cite{diehl2022distributed} and initially stored in the OTF2~\cite{eschweiler2012open} format. For configurations based on DuckDB, JSON-formatted equivalents of each trace were generated to support structured querying and flexible data manipulation. The ``Synthesized'' dataset was created by cloning the K-means dataset to increase the time span, number of tracks, and overall number events.

\begin{table}[t]
\centering
\caption{Characteristics of datasets used in the evaluation}
\label{tab:datasets}
\begin{tabular}{|l|c|c|c|}
\hline
\textbf{Dataset} & \textbf{Distribution} & \textbf{\# Tracks} & \textbf{\# of Events} \\
\hline
DGEMM           & Clustered   & 16   & 1.3K \\
K-Means         & Sparse      & 16   & 29.4K \\
LULESH          & Dense       & 49   & 160K \\
LRA & Sparse & 160 & 1.1M \\
Fibonacci       & Dense       & 8    & 2.3M \\
Synthesized     & Sparse      & 496  & 3.6M \\
\hline
\end{tabular}
\end{table}

\subsection{Queries Tested}

We evaluate two queries, one that queries all events across all tracks within a time interval (``range query'') and one that retrieves the subset of events in a time interval that meet a condition based on attribute value (``conditional range query''). These queries represent two of the most common query types used in parallel timeline charts as identified in the literature review by Sakin et al.~\cite{sakin2024gantttaxonomy}. The range query is at the heart of panning, zooming, and selecting time windows. The conditional query represents these activities under a filtered scenario and exercises ESeMan's handling of attribute data.

For the range query, we randomly generate twenty non-overlapping time ranges for each dataset, using seeds to ensure consistency across data management configurations. For conditional range queries, we selected a single categorical attribute and value per dataset such that there exists at least one event in each of our tested ranges.

\subsection{Metrics}

For each query, we measured the \textbf{Data Fetch Time}, which is the duration from the moment a query is initiated to the moment the backend returns the result to the frontend. This metric reflects the efficiency of data retrieval and query execution mechanisms. We also captured the final memory consumption using the Linux \texttt{pmap} command after completing all the query executions for each condition. 

To assess accuracy, we measured the \textbf{Structural Similarity Index Measure (SSIM)}, which reflects similarity between two images. SSIM considers changes in the structural information to better align with human visual perception~\cite{wang2004image}. A value of 0 indicates no similarity, while a value of 1 indicates a perfect match. To this end, we generate PNG~\cite{roelofs1999png} files for each of our trials and compare them to the PNG generated by the \texttt{Naive} configuration which gathers and visualizes all events.

\begin{figure*}[ht]
    \centering
    \begin{subfigure}[t]{\columnwidth}
        \includegraphics[width=\columnwidth]{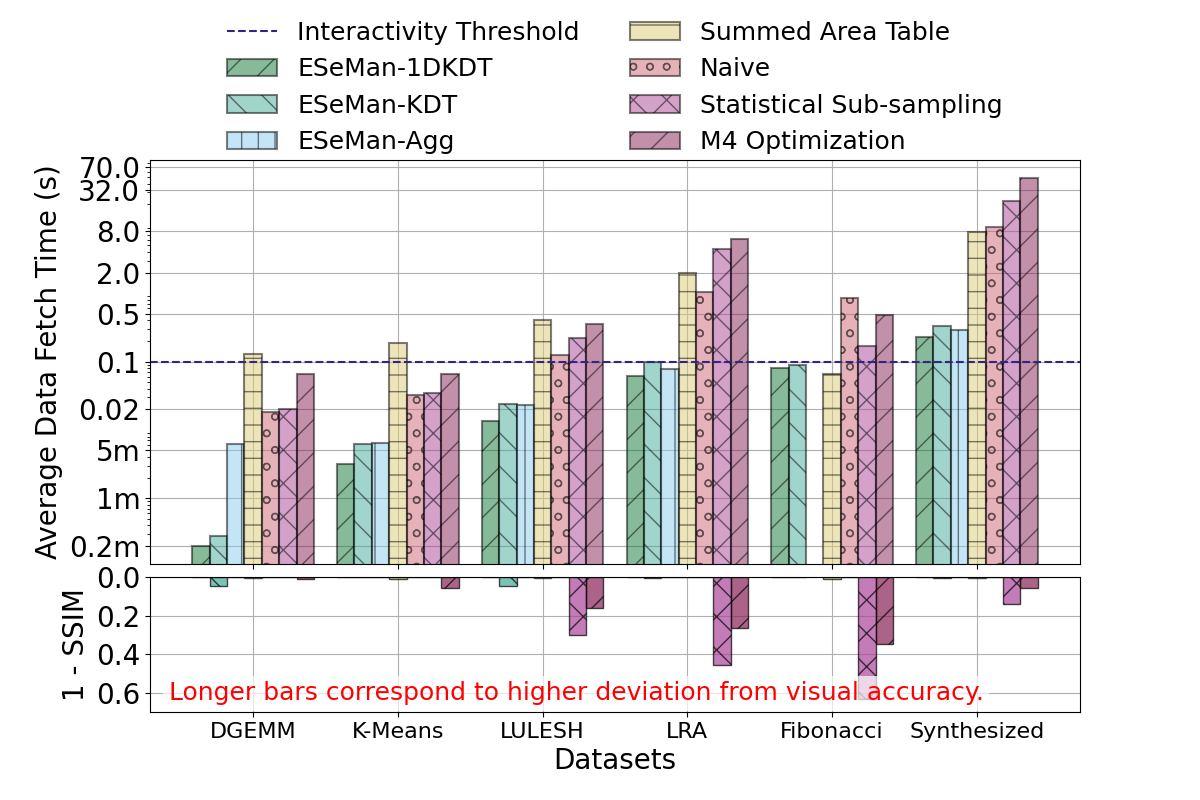}
        \caption{Range Query Performance}
    \end{subfigure}
    \hfill
    \begin{subfigure}[t]{\columnwidth}
        \includegraphics[width=\columnwidth]{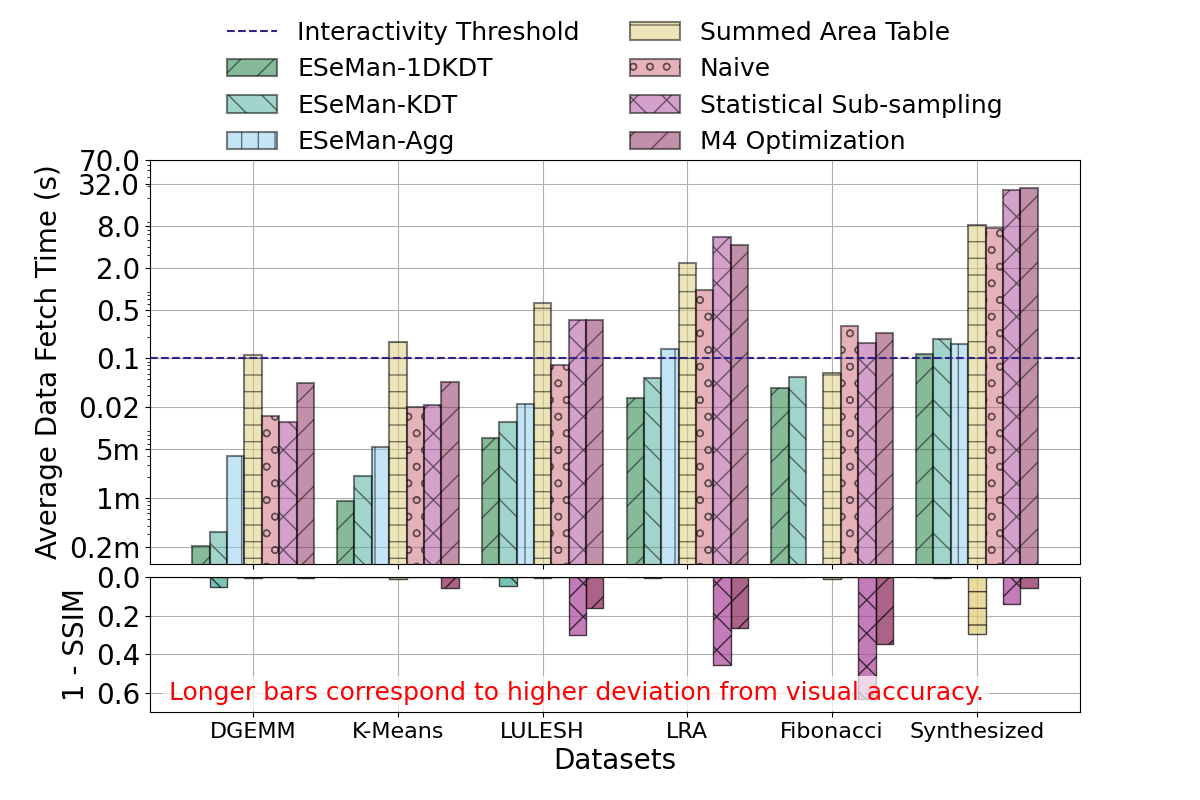}
        \caption{Conditional Range Query Performance}
    \end{subfigure}
    \caption{Fetch performance and visual accuracy comparison of seven data management configurations for the (a) range queries and (b) conditional range queries across the tested datasets. The top charts present the average data fetch time. The blue dashed line indicates the interactivity threshold (100 ms). The bottom charts show the deviation from the visual accuracy compared to the \texttt{Naive} configuration presented as (1-SSIM). ESeMan configurations achieve data fetch time within the interactivity threshold, except for the Synthesized dataset, with the highest visual accuracy.}
    \label{fig:rangeq}
\end{figure*}

\subsection{Automated Evaluation Framework}
\label{subsec:eval_framework}

We developed an automated evaluation framework to ensure consistency, repeatability, and timing accuracy in performance measurements. It uses a Selenium-based automated simulator to load the Traveler visualization interface on a Chromium web browser, using each data management configuration in turn to automate the execution of all experiment workflows.

The simulator automates key interactions within the interface, including dataset loading, initial chart rendering, and the execution of both range and conditional range queries. To emulate user-driven exploration in a reproducible way, the simulator selects interaction parameters randomly from predefined bounds using a fixed initial seed value for repetition between our conditions. 

Queries are executed twenty times each and the average of the final ten executions is measured. We run the first ten queries without measuring them to eliminate noise caused by unrelated visualization components triggered by interactions, system initialization, background processes, or first-time access delays. This choice puts the focus on steady-state behavior. 

\subsection{Hardware Configuration}

All experiments were performed on a high-performance computing cluster with an AMD EPYC 9755 CPU clocked at 2.70GHz. Although each cluster node has 1TB of available RAM, our batch script for the experiments limits 32GB of RAM during each run. This system is maintained by our university's high performance computing (HPC) center. We chose it as a representative of clusters that may be available to the audience of people who use parallel timeline charts.

\subsection{Results}
\label{subsec:results}

We present the results of our performance evaluation. We also demonstrate the trade-off between fetch time and accuracy by varying the pixel window size in ESeMan-1DKDT in \autoref{subsec:tradeoff}. Please see the supplemental materials for the time required to index each dataset and the initial chart rendering time.

\subsubsection{Range Query Performance}

\autoref{fig:rangeq}(a) presents the average data fetch time (top) and visual accuracy (reported as \(1 - \text{SSIM}\)) (bottom) across the different configurations and datasets. The ESeMan variants (\texttt{1DKDT} and \texttt{KDT}, are below the interactivity threshold of 100 ms recommended by Battle et al.~\cite{battle2019role} for all conditions except the synthesized dataset where the data fetch takes up to 0.4 seconds. The \texttt{Agg} variant exhibited similar behavior, except for on the Fibonacci dataset where it crashed due to a memory allocation error. The Summed Area Table exceeds the threshold for most datasets while  \texttt{Naive}, \texttt{Statistical Sub-sampling}, and \texttt{M4 Optimization} exceed the threshold four all but the smallest two datasets. 

In terms of visual accuracy, ESeMan exhibits high structural similarity to the \texttt{Naive} configuration, with the per-track hierarchical structures exhibiting perfect accuracy. In contrast, Sub-Sampling and M4 Optimization introduce greater visual dissimilarity. 

Among the ESeMan configurations, ESeMan-1DKDT has the lowest data fetch time. Both \texttt{ESeMan-1DKDT} and \texttt{ESeMan-Agg} have the highest visual accuracy. Comparatively, \texttt{ESeMan-KDT} has slightly less visual accuracy. Overall, the results demonstrate that ESeMan's hierarchical structure achieves both fast performance and high visual accuracy across all evaluated datasets for the range query.

\subsubsection{Conditional Range Query Performance}

\autoref{fig:rangeq}(b) presents the performance of all configurations on the conditional range query, evaluating both average data fetch time (top) and visual accuracy via \(1 - \text{SSIM}\) (bottom). Similar to the unfiltered case, ESeMan configurations (\texttt{KDT}, \texttt{2DKDT}, and \texttt{Agg}) consistently perform within the 100ms interactivity threshold across all datasets, while the other configurations often exceed the latency threshold, in some cases by an order of magnitude. 

The visual accuracy plot indicates that ESeMan maintains high structural similarity with the original data (\(1 - \text{SSIM} \approx 0\)), while sub-sampling and approximate techniques introduce greater visual distortion. These results confirm that ESeMan not only supports fast conditional querying but also preserves accuracy in the filtered visual output.

\subsubsection{Comparison of Memory Consumption}

\begin{figure}[ht]
    \centering
    \includegraphics[width=\columnwidth]{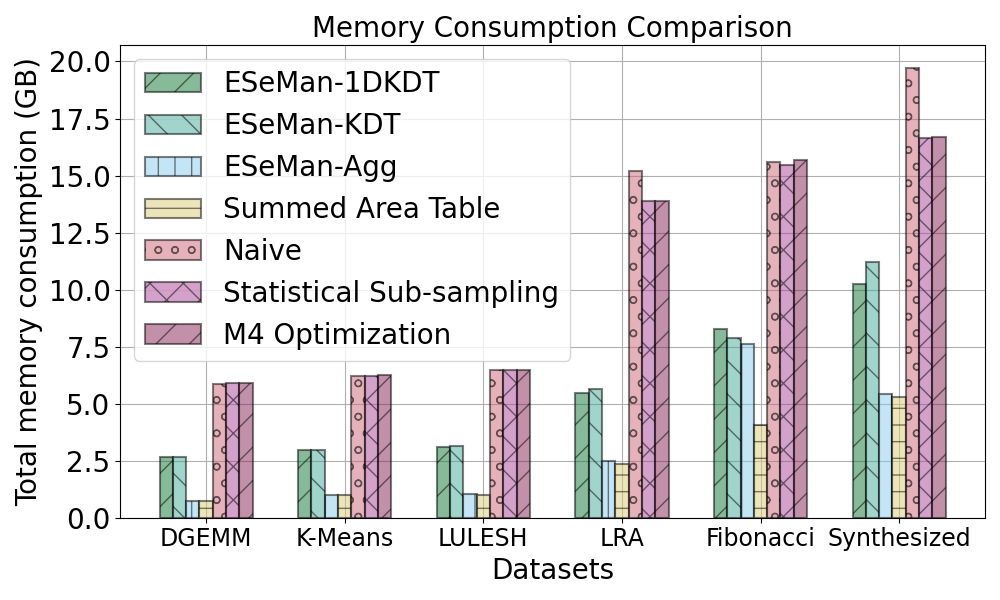}
    \caption{Total memory consumption of different configurations across datasets. The ESeMan-Agg and summed area table configurations have consistently low consumption compared to others.}
    \label{fig:memory}
\end{figure}

\autoref{fig:memory} shows the total memory consumption of different configurations across all datasets. All ESeMan configurations consume very low memory compared to DucKDB-based configurations. DuckDB-based configurations have mostly similar memory consumption, with the \texttt{Naive} being the highest across all configurations. The \texttt{Summed Area Table} has the lowest memory consumption. These results show that, for generating summaries, a linear data structure consumes the lowest memory.

LMDB, the memory-mapped database used in our ESeMan configures, requires a fixed amount of memory to be reserved. To meet this requirement,  we fixed 2GB across all of our experiments, though we may have been able achieve lower memory had we tuned our configurations per dataset. Note that the limit did not contribute to the \texttt{Agg} configuration failing on the Fibonacci dataset as that configuration does not use LMDB.

\subsubsection{Visual accuracy and latency of ESeMan-1DKDT}
\label{subsec:tradeoff}

\begin{figure}[t]
    \centering
    \includegraphics[width=\columnwidth]{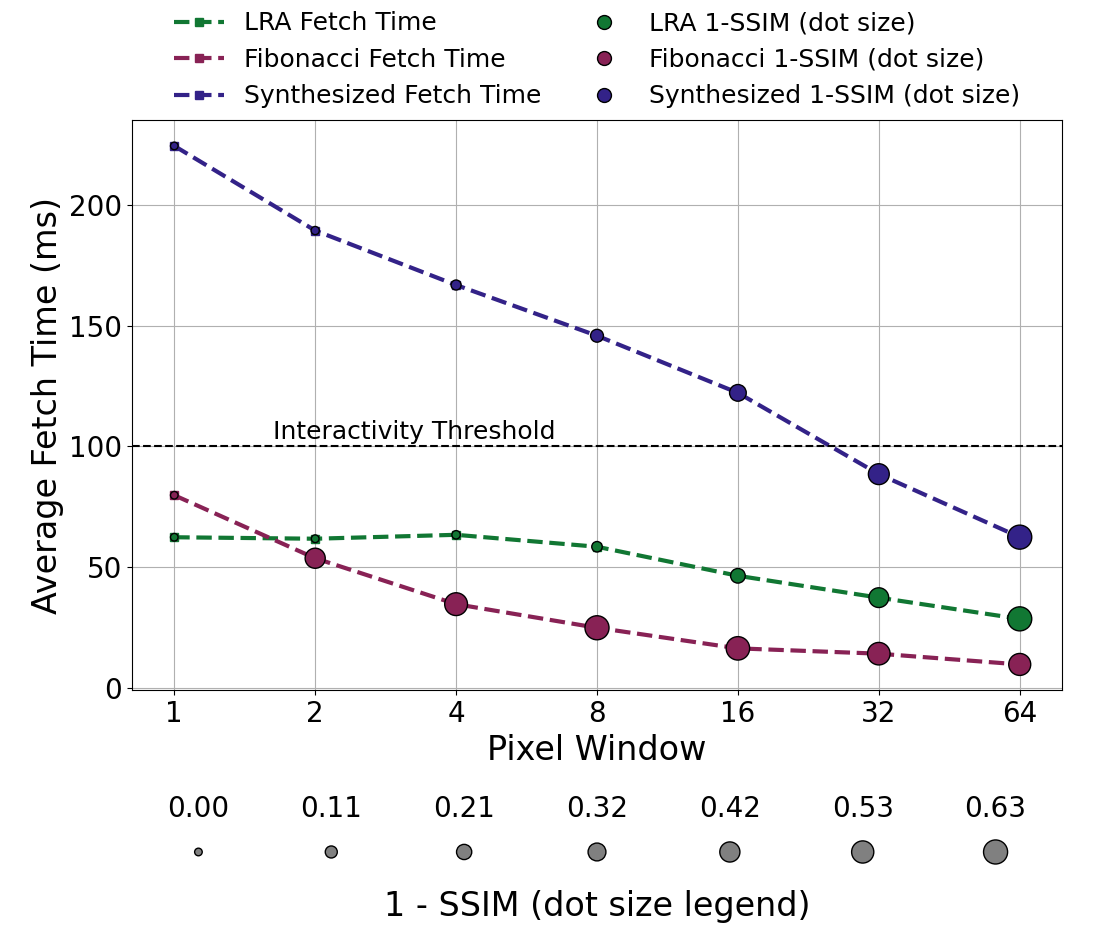}
    \caption{Comparison of \textbf{ESeMan-1DKDT} across the three largest test datasets using different pixel windows. As we increase the summarization window, the data fetch time decreases and SSIM declines, demonstrating how trade-off between interactivity and visual accuracy.}
    \label{fig:ssim}
\end{figure}

\autoref{fig:ssim} illustrates the trade-off between data fetch time and visual accuracy (measured by SSIM) for the \textbf{ESeMan-1DKDT} configuration across the three largest datasets.\footnote{The three smaller datasets had much smaller fetch times and would typically not need tuning, but exhibit similar behavior. We have included their results in the supplemental material.} We vary the pixel window used by ESeMan which causes the fetch time and accuracy to decrease. This effect is especially notable in datasets with high density per track, such as \textit{Fibonacci}, which exhibits a steep drop fetch time, but also a significantly declining SSIM value. These results showcase how adjusting the pixel window size in \textbf{ESeMan-1DKDT} allows users to tune the balance between speed and visual accuracy effectively.

\autoref{fig:teasers} shows the resulting parallel timeline charts for the LRA dataset with different pixel window configurations versus the \texttt{Naive} ground truth. When the pixel window is 1, ESeMan leads to a chart that is visually indistinguishable from the raw version (SSIM = 1.0), while reducing the fetch time to 53 milliseconds in comparison to the 499 ms needed by the \texttt{Naive} configuration. When the pixel window is further widened, the fetch time decreases but visual accuracy decreases.

\subsection{Comparison with Perfetto}
\label{subsec:perfetto}

We also compare ESeMan to \emph{Perfetto}~\cite{Perfetto}, a trace visualization tool developed by Google. We did not include Perfetto in our earlier evaluation, as it is tightly coupled with its own execution environment and data collection pipeline, making it infeasible to run experiments under consistent and comparable conditions.

To demonstrate the difference between ESeMan and Perfetto's summarization approaches, we use the K-means dataset, which has a high density of short-duration events. This dataset was specifically chosen to evaluate whether the summarization approach can preserve fine-grained temporal patterns, particularly at lower resolutions or zoomed-out views. Maintaining the visibility of short events under visual summarization is required for visual anomaly detection tasks during visual exploratory analysis~\cite{guo2021survey}.

We loaded the dataset into both Perfetto and ESeMan and captured a snapshot of the generated visualization. Perfetto employs a high-level summarization approach that selectively omits data, particularly at lower zoom levels, in order to optimize rendering speed and interactivity. While this leads to faster response times, it comes at the cost of reduced detail and potential loss of important visual patterns in the trace data. 

In contrast, ESeMan preserves pixel-level accuracy throughout the interaction, rendering all events without omission, even at fine temporal resolutions, providing users with a more faithful representation of the trace. 

ESeMan maintained a competitive interaction latency to Perfetto even with all the events shown. We compared the \emph{interaction-to-next-point} (INP~\cite{inp_webdev}) values of the two systems. INP is a web performance metric that measures the latency between user input and the visual response. Specifically, ESeMan-1DKDT demonstrated an INP value of 23 ms, which is comparable to Perfetto's 20 ms. Relevant data has been provided in the supplemental materials.

\subsection{Limitations and Threats to Validity}
Though the experimental setup highlights the efficiency and accuracy of our proposed solutions, there are several limitations to consider. First, while we used datasets from real applications~\cite{sakin2022traveler} and ensured varying data characteristics, they all come from a single domain. Other domains, such as medical, finance, or project management may task ESeMan in different ways.

Although we accounted for caching behavior by averaging over repeated queries, real-time systems may exhibit more variable performance due to contention or concurrent tasks.

Queries executed over DuckDB were not heavily optimized. It is possible that further research in optimizing them for parallel timeline charts could close the performance gap with our indexing-based solutions.

Another limitation stems from our choice of query types. While range and conditional range queries are representative of common visual exploration tasks~\cite{sakin2024gantttaxonomy}, visualization tasks related to presenting attribute details, highlighting dependencies among the events, and highlighting patterns were not explicitly tested. Additionally, evaluating summarization quality under task-driven scenarios (e.g., anomaly detection, root cause analysis) would provide a more application-aligned perspective on performance.

\section{Discussion}
\label{sec:discussion}

Our results demonstrated that ESeMan was the most efficient data management scheme for parallel timeline charts while also maintaining near perfect accuracy with respect to the naive approach, outperforming other popular data management options. While the results were consistently strong across all of our ESeMan configurations, users should choose a configuration that suits their tasks. As the one-dimensional KD-tree configuration was the most efficient, robust, and accurate, we recommend it be used in most circumstances. However, when memory consumption is a limiting factor, the agglomerative clustering configuration can be used as it typically requires half the memory of the KD-tree version while maintaining similar accuracy and fetch times for larger datasets.

While we expected the 2D KD-tree configuration to outperform the 1D case when given a fixed track order, it was typically slightly slower than the 1D case. Further analysis revealed that, compared to the 1D case, the 2D KD-tree has a larger tree height (farthest distance from the root node to any leaf node) and higher total node counts, which led to unnecessary node traversal during each query of our experiment. The higher node counts also resulted in increased data lookups in LMDB which further decreased the potential efficiency gains. Additional investigation is necessary to explore different splitting rules (for example, consecutive splits in the time dimension) and tree traversal techniques (for example, utilizing a priority queue based on the splitting rule) to reduce unnecessary node visits, and ultimately, identify which approach yields better query performance.

ESeMan could be configured with a different database system. As noted in our Limitations section, DuckDB queries could have been further optimized as well.
In database-only approaches, query performance is highly dependent on index design, SQL formulation, and execution planning. Our results suggest that achieving interactive latency in such systems often requires careful, manual optimization, placing a significant burden on visualization designers. With \textbf{ESeMan}, we attempt to alleviate this burden by providing a purpose-built, indexing-aware data manager that ensures low-latency interaction without requiring deep expertise in database tuning.

While our current evaluation focuses on query performance on static datasets, scenarios with a dynamic dataset where existing event data changes or new events arrive after data ingestion require further investigation. Additionally, we conducted our experiments in single-node cases, capping the dataset size due to memory limitations. Scaling further would require updating the implementation and testing ESeMan using distributed components.

ESeMan represents a significant step forward in scalable event sequence visualization, demonstrating the possibility of achieving both interactive performance and visual accuracy in large exploratory data exploration. By addressing the fundamental trade-offs that have limited previous approaches, our work opens new possibilities for effective interactive analysis of large event sequences across diverse application domains.

\section{Conclusion}
We introduced ESeMan (Event SEquence MANager), a hierarchical data management system designed to address the dual challenges of maintaining interactive performance and visual accuracy in large event sequence visualization using parallel timeline charts. ESeMan demonstrates significant improvements over popular existing approaches in both data retrieval performance and visual accuracy. Our presented evaluation validates ESeMan's effectiveness in achieving sub-100 ms interactive response times while maintaining pixel-perfect visual accuracy. Furthermore, we show how ESeMan's tunable pixel window allows trading off mild visual accuracy for faster data fetch times, thereby increasing the effective data scale at which ESeMan can serve interactive queries.

\acknowledgments{
 We thank Dr. Alexander Lex and Dr. Paul Rosen for their suggestions and feedback on the library design and the evaluation. We thank Dr. Kevin A. Huck for providing several of the datasets we used.

 This work was supported by the Department of Energy under DE-SC0022044/DE-SC0024635 and the National Science Foundation under NSF IIS-1844573/IIS-2324465. 
}

\bibliographystyle{abbrv-doi}

\bibliography{main}
\end{document}



\firstsection{Guide to Supplemental Materials}

\maketitle

This document contains a description of all supplemental materials.

The ESeMan library can be downloaded from this git repository: \href{https://github.com/sayefsakin/eseman}{https://github.com/sayefsakin/eseman}.

All the supplemental materials are provided in this \href{https://osf.io/vdcwx/?view_only=99748e7f1ac14886814ffb5404e82c60}{OSF Link}.

All materials are included at the above-linked OSF .zip file as:
\begin{itemize}
    \item plot\_scripts: This folder contains all scripts we utilized to plot the results in the paper.
    \item perfetto\_tests: This folder has all relevant files for the case study we conducted using Google's Perfetto.
    \item source\_code: All source code files for our tested visualization tool and benchmark harness framework.
    \item batch\_scripts: All source code files that we used to run our experiment on a HPC cluster.
    \item ecc21d0a-112a-4b52-8cdd-6aca80adde93: Results folder for the DGEMM dataset.
    \item a01b2607-32a6-4435-a2ae-20a4d227e5fd: Results folder for the LULESH dataset.
    \item f4e2fdfa-893e-4f13-bac8-e9fbbdf40c1f: Results folder for the LRA dataset.
    \item faf17535-2f66-4621-995f-49c7dbd84e8b: Results folder for the K-Means dataset.
    \item ae63b22a-66a3-4e92-ae49-b8206d8a0e7b: Results folder for the Fibonacci dataset.
    \item 908fc737-2cc7-41d8-8281-7dd9e83155ff: Results folder for the Synthesized dataset.
    \item (pixel-window)\_(dataset-id): Results folder for the figure 8 presented in the main paper.
\end{itemize}

All file names in each results folder follow the following pattern,
\begin{itemize}
    \item figures: This folder contains all the generated figures during the experimentation that we used to calculate SSIM.
    \item (dataset-name)\_(query)\_merged.csv: This file contains the query execution results
    \item (dataset-name)\_(query)\_selenium\_check: This file contains the INP result by the selenium chrome driver.
    \item (dataset-name)\_(query)\_cgal\_check: This file contains the query execution results by the ESeMan configurations.
\end{itemize}

Our actual raw dataset and indexed dataset sizes are very large; therefore, they have not been included in the supplemental material.

\section{DuckDB configurations}
In DuckDB, we created a table $intervals$ with all the event sequences. Then we conduct the following queries over the specified range $(begin, end)$ on the track $Location$ with a predefined number of $bins$, which is set to the visible number of pixels $(4236)$.
\paragraph{Naive Range Query}
This query selects raw trace entries within a specified time range without any optimization.
\begin{lstlisting}[language=SQL,caption={C3 - Naive Range Query}]
SELECT enter_timestamp, leave_timestamp, Location 
FROM intervals
WHERE Location = location
    AND leave_timestamp >= begin
    AND enter_timestamp <= end
\end{lstlisting}

\paragraph{Statistical Sub-sampling}
Here, we use reservoir sampling to reduce data volume while preserving statistical characteristics.
\begin{lstlisting}[language=SQL,caption={C4 - Statistical Sub-sampling Query}]
SELECT enter_timestamp, leave_timestamp, Location
FROM intervals
WHERE Location = location
    AND leave_timestamp >= begin
    AND enter_timestamp <= end
USING SAMPLE reservoir(bins ROWS) REPEATABLE(100);
\end{lstlisting}

\paragraph{M4 Optimization}
This query uses aggregation over fixed-size time bins (M4 optimization), returning representative min/max pairs.

\begin{lstlisting}[language=SQL,caption={C5 - M4 Optimization Query}]
WITH Q AS (
    SELECT enter_timestamp as atime, 1 AS ct 
    FROM intervals
    WHERE leave_timestamp >= begin 
        AND enter_timestamp <= end 
        AND Location = location
    UNION
    SELECT leave_timestamp as atime, 0 AS ct 
    FROM intervals
    WHERE leave_timestamp >= begin 
        AND enter_timestamp <= end 
        AND Location = location
    ORDER BY atime
)
SELECT k, atime, Q.ct 
FROM Q JOIN (
    SELECT round(bins*(atime - begin)/(end - begin))AS k, 
        min(atime) as min_atime,
        max(atime) as max_atime
    FROM Q GROUP BY k
) as QA ON k = round(bins*(atime - begin)/(end - begin))
    AND (atime = min_atime or atime = max_atime) 
ORDER BY k, atime
\end{lstlisting}

\begin{table*}[t]
\centering
\caption[]{Indexing time and indexed file size for each dataset across data management configurations.\footnotemark}
\label{tab:indexing-results}
\small
\renewcommand{\arraystretch}{1.2}
\setlength{\tabcolsep}{6pt}
\begin{tabular}{|l|c|c|c|c|c|c|c|c|}
\hline
\multirow{2}{*}{\textbf{ }} & 
\multicolumn{4}{c|}{\textbf{Indexing Time}} & 
\multicolumn{4}{c|}{\textbf{Database Size}} \\
\cline{2-9}
& \textbf{ESeMan-1DKDT} & \textbf{ESeMan-KDT}& \textbf{Summed Area Table} & \textbf{DuckDB} 
& \textbf{ESeMan-1DKDT} & \textbf{ESeMan-KDT}& \textbf{Summed Area Table} & \textbf{DuckDB} \\
\hline
DGEMM & 0.7s & 0.8s & 0.2s & 0.2s & 0.6M & 0.5M & 0.9M & 3M \\
K-Means & 5.9s & 8s & 2.4s & 3.2s & 15M & 9M & 35M & 8M \\
LULESH & 47.8s & 70s & 80s & 157s & 63M & 64M & 50M & 50M \\
LRA & 652.8s & 315s & 504s & 147.5s & 357M & 327M & 213M & 235M \\
Fibonacci & 788.1s & 1129s & 196.5s & 251.3s & 2.6G & 1.2G & 453M & 482M \\
Synthesized & 2971.5s & 1113s & - & - & 1.1G & 1.3G & 674M & 650M \\
\hline
\end{tabular}
\end{table*}

\section{Dataset Indexing Time}

Dataset indexing refers to the time required to ingest and preprocess trace data before it becomes available for interactive use within the Traveler ecosystem. This step includes parsing the trace file, extracting event metadata, and storing it in the appropriate backend (LMDB, MySQL, or DuckDB), depending on the configuration.

\autoref{tab:indexing-results} \footnotetext{DuckDB-based configurations share identical indexing pipelines, including the JSON conversion, and thus report the same values. ESeMan-Agg is in-memory; therefore, it is not reported here. Synthesized data followed a different pipeline to construct the Summed area table and DuckDB. Therefore, its construction time is also not reported here as it will not be a fair comparison.} presents the dataset indexing time and total dataset size. For the dataset indexing time, ESeMan-based configurations do have a higher construction time compared to the other configurations for the larger datasets. Similarly, the indexed dataset size is low with the smaller datasets for the ESeMan configurations compared to others, and the dataset size is larger for the large dataset sizes compared to the other configurations.

\section{Initial Data Fetch Time}

Initial data fetch time measures how long it takes to fetch the data for the first visual view of a trace once it is loaded. The initial rendering time reported in our experiments does not include the startup time of the visualization server itself. Instead, it is measured from the point after the Chromium browser is launched and the visualization interface is loaded via its URL. This timing captures only the duration required for the client-side application to request and render the initial view after the server is already running.

\begin{figure}[h]
    \centering
    \includegraphics[width=\columnwidth]{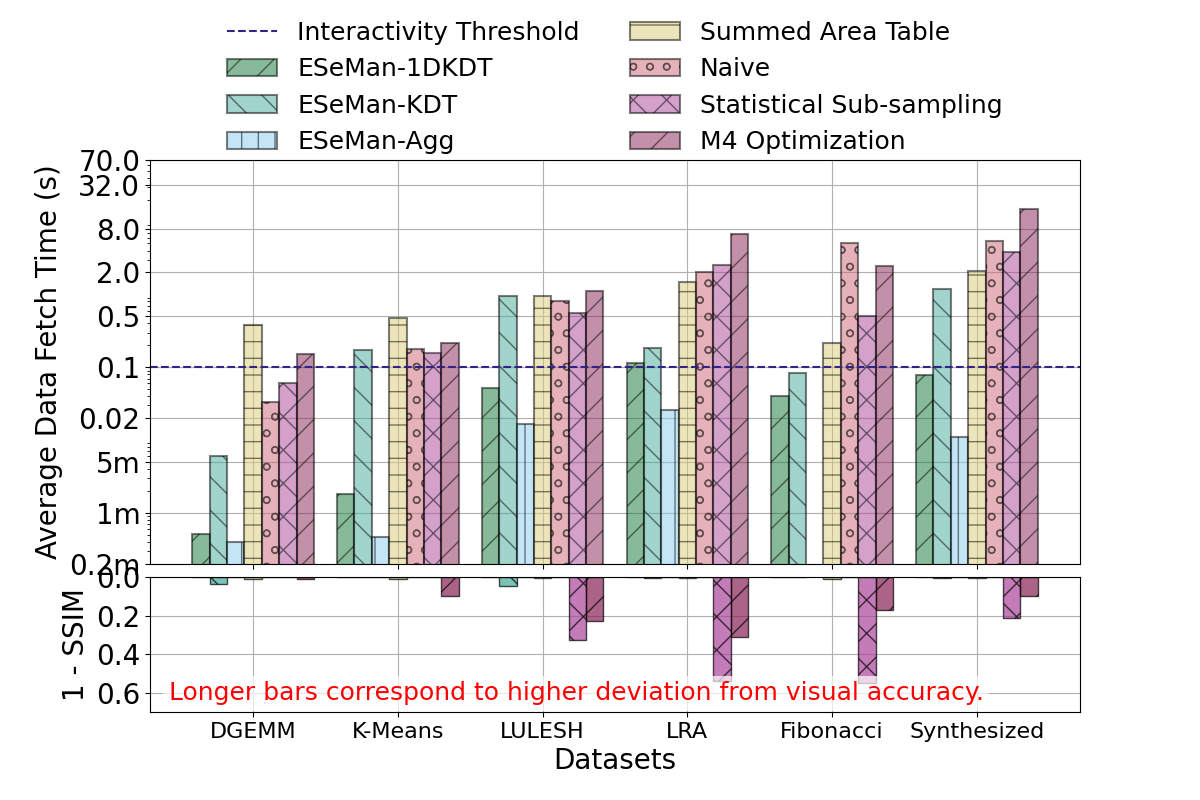}
    \caption{Initial chart rendering time across datasets}
    \label{fig:initial-render-time}
\end{figure}

\autoref{fig:initial-render-time} presents the data fetch time for the initial render times. Although for the smaller datasets DGEMM and K-Means, the data fetch time is within the interactivity threshold limit of 100 milliseconds, all configurations take a very long time to render the visualization with a larger dataset. However, we notice ESeMan-Agg constantly requires less time to load across all datasets. Due to the sparse distribution of data, ESeMan-KDT takes comparatively longer time with the Synthesized dataset. For visual accuracy, ESeMan-1DKDT and ESeMan-Agg consistently performed best compared to others across all datasets. The statistical subsampling has the worst accuracy among the configurations.

\begin{table}[t]
\centering
\caption{SSIM comparison across different configurations visualizing D2. SSIM values are calculated considering C3 as the baseline. Pixel-wise differences are highlighted using a colormap.}
\label{tab:ssim-results}
\renewcommand{\arraystretch}{1.2}
\begin{tabular}{|c|c|}
\hline
\multirow{2}*{\textbf{Generated Visualization}} & \textbf{Configuration} \\ & (SSIM) \\
\hline
\multirow{4}*{\includegraphics[width=0.6\columnwidth]{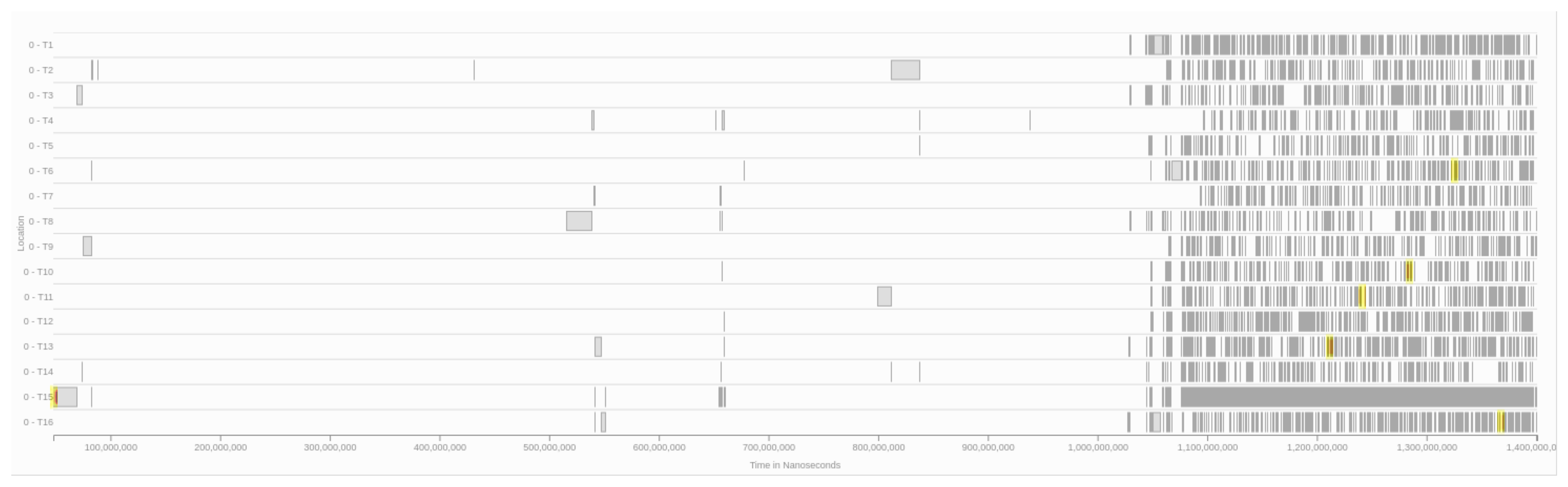}} & Summed Area Table \\ & (0.9996) \\ & \\ & \\ \hline
\multirow{4}*{\includegraphics[width=0.6\columnwidth]{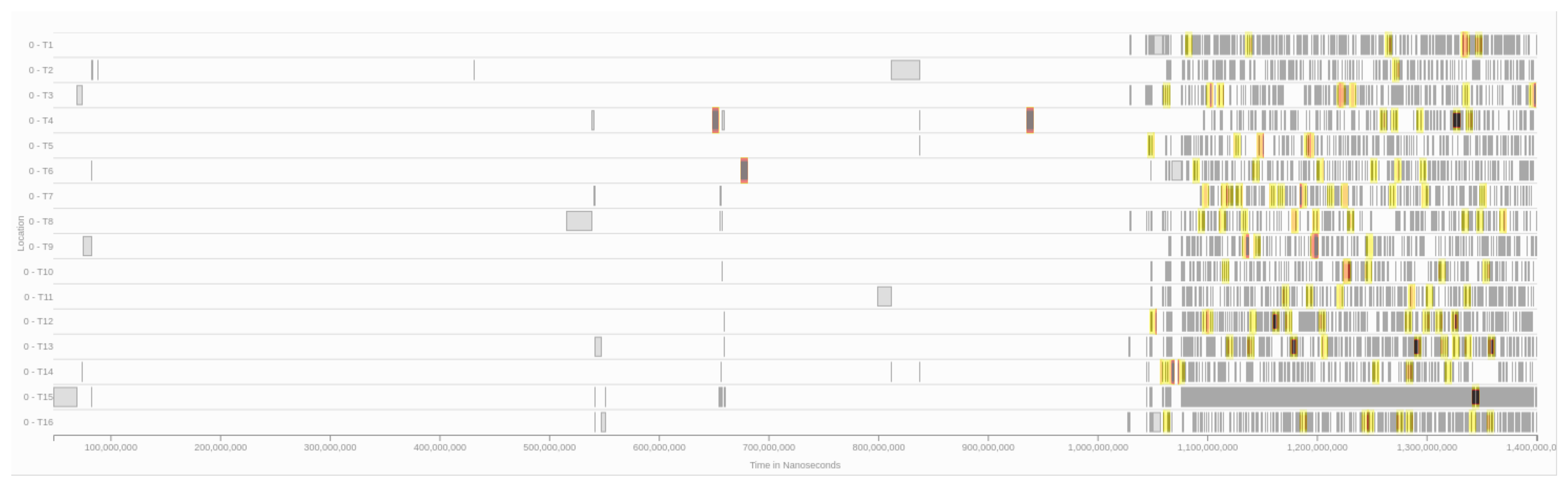}} & Statistical subsampling \\ & (0.9915) \\ & \\ & \\ \hline
\multirow{4}*{\includegraphics[width=0.6\columnwidth]{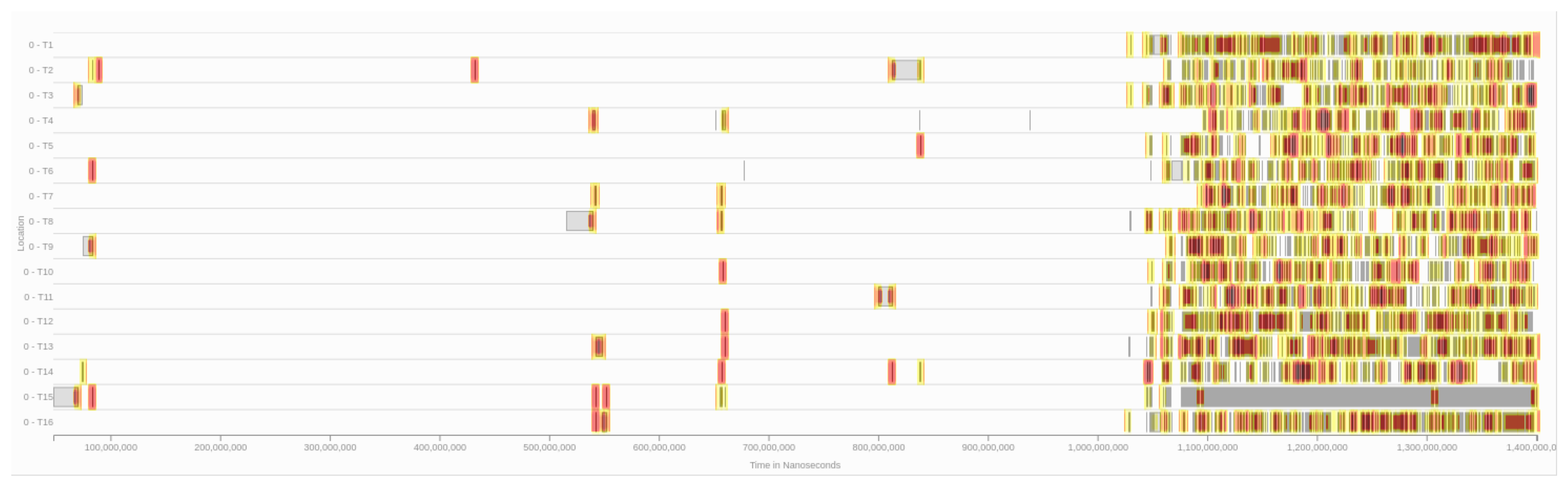}} & M4 Optimization \\ & (0.8908) \\ & \\ & \\ \hline
\multicolumn{2}{c}{\includegraphics[width=0.9\columnwidth]{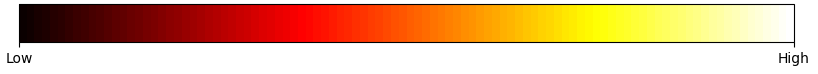}} \\
\end{tabular}
\end{table}

\section{Summarization Quality Analysis}

To demonstrate the visual accuracy using SSIM, we present here the generated visualization with all configurations using the K-Means dataset and highlight the differences compared to the visualization generated by the Naive configuration. SSIM provides an objective metric for assessing image quality that correlates well with human visual perception by considering luminance, contrast, and structural information. We considered the DuckDB Naive configuration as the baseline ground truth, as it collects all data within the range and draws the events. ESeMan-1DKDT, ESeman-KDT, and ESeMan-Agg have the perfect similarity score (SSIM=1.0). We compare the other configuration with the Naive, present the visual differences, and reporte the SSIM scores.

The visualization accuracy is fundamentally determined by the generating summary with the pixel window strategy, where the total display width is divided into a fixed number of pixel window for data summarization. When the number of pixel windows equals the number of horizontal pixels in the visualization, each bin corresponds to exactly one pixel, representing the theoretically ideal pixel-perfect rendering. As the number of pixel windows decreases below the pixel count, each pixel window must represent multiple pixels, leading to spatial summarization that inherently reduces visual accuracy. This summarization relationship with the pixel window directly impacts the effectiveness of different data management approaches, as some methods better preserve fine-grained spatial information than others when operating under reduced pixel window constraints.

The visual accuracy assessment results presented in~\autoref{tab:ssim-results} show that the DuckDB and summed area table configurations show varying degrees of quality degradation. Here, we highlight the SSIM values in using a colormap presented at the bottom of ~\autoref{tab:ssim-results}. Darker pixels represent low SSIM values (high deviation from the Naive configuration), and lighter pixels represent high SSIM values (high similarity with the Naive configuration). The Summed Area Table approach maintains high accuracy with an SSIM of 0.9996, while the Statistical Sub-sampling method shows moderate quality loss at 0.9915. The M4 Optimization technique exhibits the most significant visual degradation with an SSIM of 0.8908, as evidenced by the visible artifacts and high color distortions in the corresponding visualization. We suspect the M4 optimization is optimized for time series data only, and for event sequences, it omits consecutive longer events. These results confirm that ESeMan's KD-Tree-based aggregation strategy preserves visual integrity while maintaining computational efficiency, outperforming both statistical approximation methods and traditional database querying approaches in terms of visualization quality.


\bibliographystyle{abbrv-doi}
